\documentclass{appolb}
\usepackage{graphicx}
\usepackage{epsfig}

\begin{document}
\title{OBSERVATION OF GROUND-STATE\\TWO-NEUTRON DECAY
\thanks{Presented at the Zakopane School on Nuclear Physics ``Extremes of the Nuclear Landscape'', August 27 - September 2, 2012, Zakopane, Poland.} }

\author{M. Thoennessen$^{a,b}$, Z. Kohley$^{a}$, A.~Spyrou$^{a,b}$, E.~Lunderberg$^c$, P.A.~DeYoung$^c$, H.~Attanayake$^d$, T.~Baumann$^a$, D.~Bazin$^a$, B.~A.~Brown$^{a,b}$ G.~Christian$^{a,b}$, D.~Divaratne$^d$, S.~M.~Grimes$^d$, A.~Haagsma$^e$, J.E.~Finck$^e$, N.~Frank$^f$, B.~Luther$^g$, S.~Mosby$^{a,b}$ T.~Nagi$^c$, G.F.~Peaslee$^c$, W.~A.~Peters$^h$, A.~Schiller$^d$, J.~K.~Smith$^{a,b}$, J.~Snyder$^{a,b}$, M.~Strongman$^{a,b}$, and A.~Volya$^i$
\address{
$^a$NSCL, Michigan State University, East Lansing, MI-48824\\
$^b$Department of Physics $\&$ Ast., Michigan State Univ., East Lansing, MI-48824\\
$^c$Department of Physics, Hope College, Holland, MI-49423\\
$^d$Department of Physics $\&$ Ast., Ohio University, Athens, OH-45701\\
$^e$Department of Physics, Central Michigan Univ., Mt.~Pleasant, MI-48859\\
$^f$Department of Physics \& Ast., Augustana College, Rock Island, IL-61201\\
$^g$Department of Physics, Concordia College, Moorhead, MN-56562\\
$^h$Department of Physics $\&$ Ast., Rutgers University, Piscataway, NJ-08854\\
$^i$Department of Physics, Florida State University, Tallahasee, FL-32306
}}
\maketitle

\begin{abstract}
Neutron decay spectroscopy has become a successful tool to explore nuclear properties of nuclei with the largest neutron-to-proton ratios. Resonances in nuclei located beyond the neutron dripline are accessible by kinematic reconstruction of the decay products. The development of two-neutron detection capabilities of the Modular Neutron Array (MoNA) at NSCL has opened up the possibility to search for unbound nuclei which decay by the emission of two neutrons. Specifically this exotic decay mode was observed in $^{16}$Be and $^{26}$O.
\end{abstract}


\section{Introduction}

Exploration of nuclear systems with the largest neutron excess requires the study of neutron-unbound states and nuclei. The first neutron-unbound nucleus was already discovered in 1937 by Williams, Shepperd, and Haxby who used the transfer reaction $^{7}$Li(d,$\alpha$)$^{5}$He to deduce the existence of $^{5}$He from the measured $\alpha$-particle spectrum \cite{Wil37}. It took almost 30 years until the next unbound resonance was unambiguously identified with the determination of the scattering length of the dineutron system in 1965 \cite{Had65}. An overview of the discovery of light neutron-unbound nuclei can be found in reference \cite{Tho12}.

\begin{figure}
\centerline{\includegraphics[height=.32\textheight]{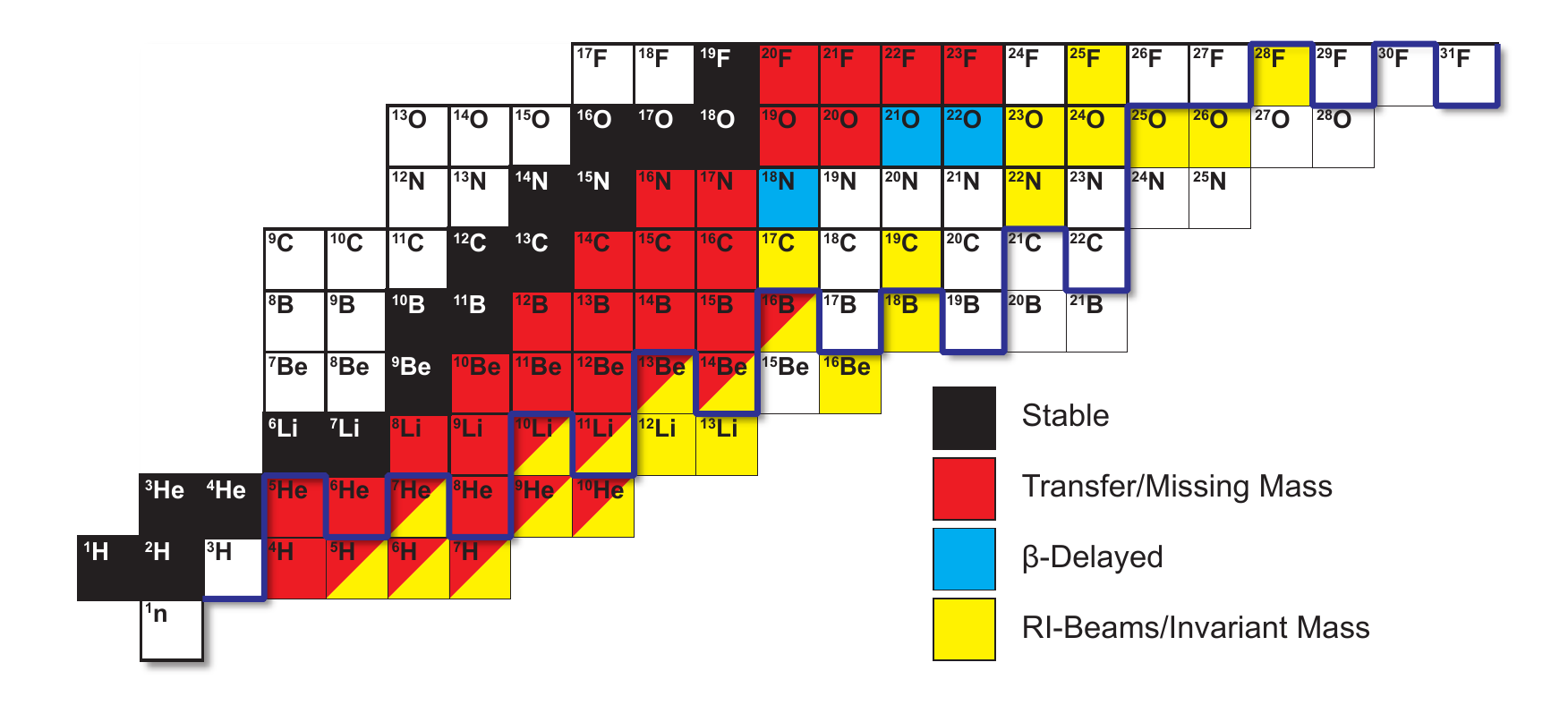}}
\caption{
Section of the chart of nuclei. Stable nuclei are indicated by black squares and the dark blue line denotes the neutron drip line.
Note that the drip line is experimentally only verified up to oxygen and the status of heavier fluorine isotopes has not been determined in
experiments. Nuclei for which neutron-unbound states have been measured are color coded according to their detection technique: missing
mass with transfer reactions (red), $\beta$-delayed neutron decay (blue) and invariant mass (yellow) measurements. The white squares beyond the
drip line indicate isotopes which have been shown to be neutron unbound. (Adapted from Ref. \cite{Bau12})}
\label{f:chart}
\end{figure}

Transfer reactions with stable beams were initially the best method to populate and study unbound states, although pion induced reactions were also an effective tool by utilizing the missing mass method. These reactions are limited to the lightest unbound nuclei where the dripline is relatively close to the stable isotopes which have to be available as targets. In order to reach heavier unbound systems new methods had to be developed. Figure \ref{f:chart} gives an overview of the different methods which were first used to observe neutron unbound states from helium to fluorine. Nuclei where the unbound states were first observed with transfer reactions with stable or pion beams dominate the region close to the valley of stability and are shown in red. Beta-delayed neutron emission was utilized to observe unbound states in $^{18}$N, $^{21}$O, and $^{22}$O for the first time (blue); this method was also used to study many of the nuclei closer to stability. Invariant mass measurements shown in yellow were necessary to populate even more neutron-rich nuclei. The diagonally shaded nuclei were explored with the missing mass as well as invariant mass method. The white squares beyond the dripline (solid thick line) show the isotopes which have been demonstrated to be unbound but no spectroscopy data has been measured. Although not explicitly stated, $^{17}$Be and $^{23}$C are almost certain to be unbound because at least one heavier isotone for these isotopes has already been shown to be unbound. Similarly, $^{18}$Be is not expected to be bound because already $^{16}$Be is unbound. Thus, all bound isotopes up to oxygen have been observed. However, there are still several bound isotopes where no unbound excited states have been observed as well as several unbound isotopes where the resonance parameters are not yet known.

The first nucleus demonstrated to be unbound with respect to two-neutron emission was $^{5}$H \cite{Gor87}. Subsequently resonances in $^{10}$He \cite{Kor94} and $^{13}$Li \cite{Aks08} were reconstructed using invariant mass measurements with radioactive beams. In the present paper we discuss results of recent two-neutron measurements of $^{16}$Be \cite{Spy12} and $^{26}$O \cite{Lun12}.

\section{$^{16}$Be}

\begin{figure}[b]
\centerline{\includegraphics[height=.27\textheight]{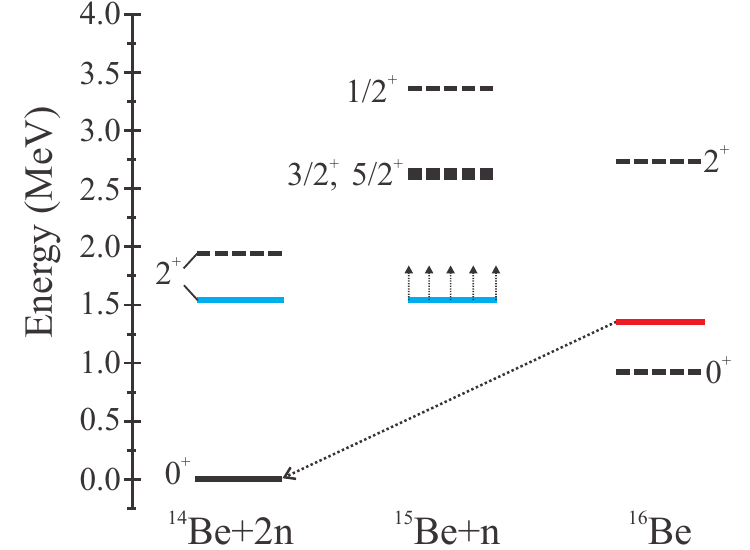}}
\caption{Level and decay scheme for neutron-rich isotopes of beryllium. The solid blue lines represent previous experimental observations of the first excited state in $^{14}$Be \cite{Sug07} and a lower limit on the position of the $^{15}$Be ground state \cite{Spy11}. The dashed black lines represent shell model calculations within the $s$-$p$-$sd$-$pf$ model space using the WBP interaction \cite{War92}. The recently measured $^{16}$Be ground state \cite{Spy12} is shown as the solid red line (Adapted from Ref. \cite{Koh12a}).}

\label{f:Be-level}
\end{figure}

$^{16}$Be was predicted to be bound with respect to one-neutron emission but unbound with respect to two-neutron emission and is thus an ideal case to search for correlated two-neutron or dineutron-like decay. In the previous cases ($^{5}$H and $^{10}$He) the intermediate unbound systems ($^{4}$H and $^{9}$He, respectively) have broad resonances, which can extend below the single-neutron emission threshold and thus favor sequential decay. The situation is predicted to be different as shown in Figure \ref{f:Be-level}. The energies levels of $^{14}$Be, $^{15}$Be and $^{16}$Be calculated with the shell model in the $s$-$p$-$sd$-$pf$ model space using the WBP interaction \cite{War92} indicate that the only open decay path is the direct emission of two neutrons. This is supported by the non-observation of $^{14}$Be in the recent two-proton removal reaction from $^{17}$C which determined the lower limit of the $^{15}$Be ground state to be at the energy of the first excited state of $^{14}$Be \cite{Spy11}.

The decay of $^{16}$Be was measured following the one-proton removal reaction from $^{17}$B at the Coupled Cyclotron Facility at NSCL/MSU. Neutrons were measured with MoNA in coincidence with charged fragments. The three-body decay energy as well as the neutron-neutron-$^{14}$Be correlations were measured \cite{Spy12}. The neutron interactions were simulated with GEANT4 \cite{Ago03} using the physics class MENATE\_R \cite{Roe08} in order to distinguish true two-neutron events from a single neutron interacting twice in the detector array \cite{Koh12b}.

\begin{figure}
\centerline{\includegraphics[height=.26\textheight]{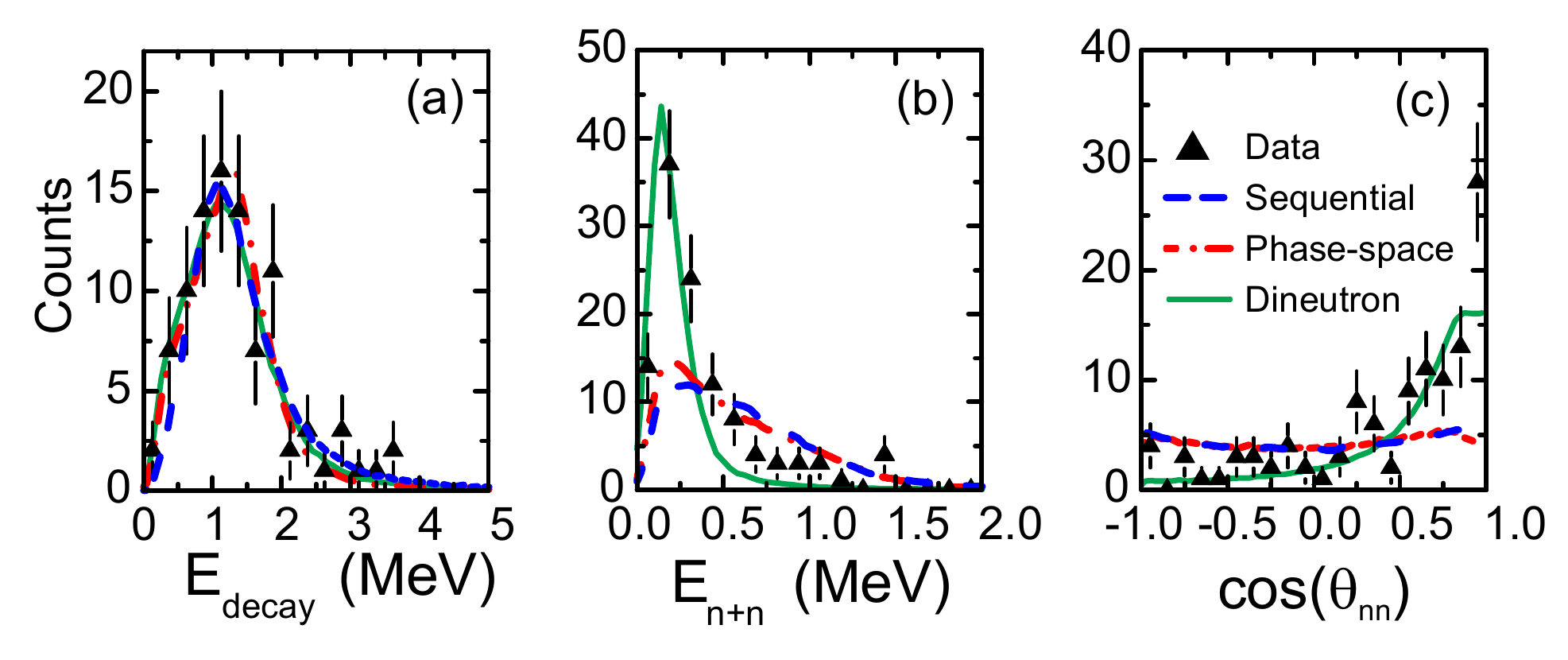}}
\caption{
(a) Three-body decay energy from the reconstruction of $^{14}$Be+n+n, (b) two-body relative energy of the two neutrons, (c)
opening angle $\theta_{\mathrm{n-n}}$ between the two neutrons in the center-of-mass frame of $^{16}$Be. The experimental data are shown in
black triangles, sequential emission in dashed lines, simultaneous phase-space emission in dot-dashed lines and dineutron decay in solid
lines (Adapted from Ref. \cite{Spy12}).}
\label{f:16Be}
\end{figure}

Figure \ref{f:16Be} shows the three-body decay energy from the reconstruction of $^{14}$Be+n+n (a), the two-body relative energy of the two neutrons (b), and the opening angle $\theta_{\mathrm{n-n}}$ between the two neutrons in the center-of-mass frame of $^{16}$Be (c). Simulations corresponding to three different decay modes were performed. Sequential emission (dashed lines) and simultaneous phase-space emission (dot-dashed lines) cannot reproduce the two-neutron relative energy nor the two-neutron opening angle. Only the dineutron decay simulation can reproduce all three spectra. It should be mentioned that the dineutron decay as simulated in a two-body model is only an approximation and full correlated three-body model calculations are necessary in order to describe the decay of $^{16}$Be more realistically.

\section{$^{26}$O}

\begin{figure}[b]
\centerline{\includegraphics[height=.55\textheight]{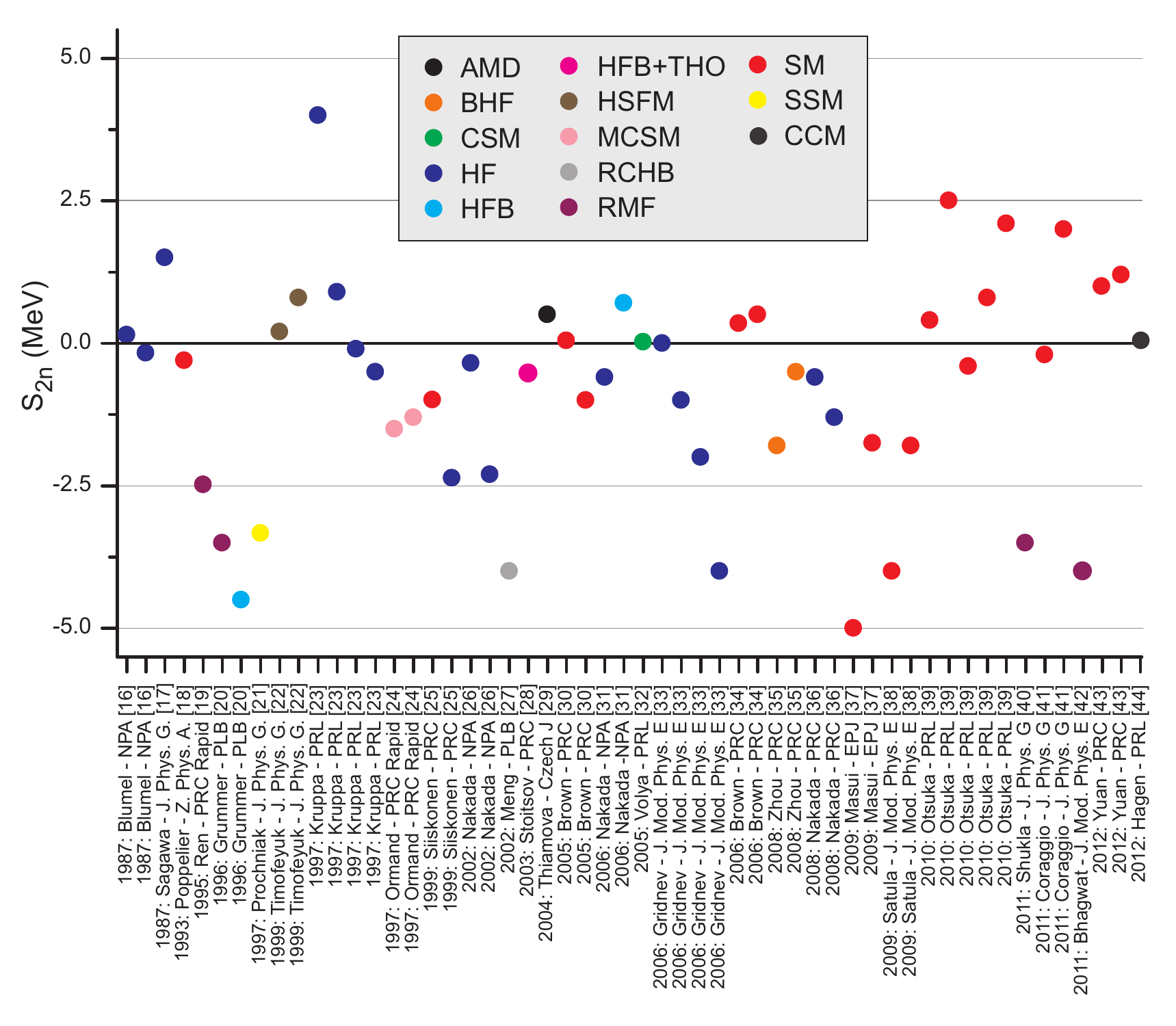}}
\caption{
Predictions for the two-neutron separation energy S$_{2n}$ for $^{26}$O from various theoretical models.}
\label{f:ox-lit}
\end{figure}

\begin{figure}
\centerline{\includegraphics[height=.3\textheight]{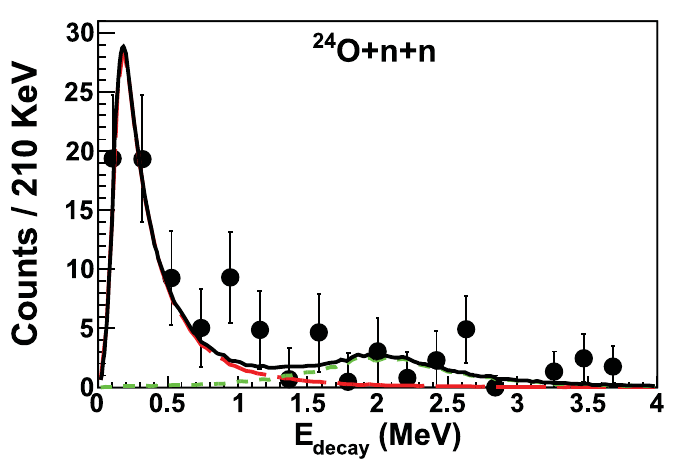}}
\caption{
Decay energy spectrum for $^{26}$O  (Adapted from \cite{Lun12}).}
\label{f:26O-data}
\end{figure}

Another example of a possible dineutron emitter is $^{26}$O. The predictions for the two-neutron separation energies show significant differences ranging from 5~MeV bound to 3~MeV unbound (see Figure \ref{f:ox-lit}). Several experimental searches for bound $^{26}$O were unsuccessful \cite{Gui90,Fau96,Tar97,Sch05}. The definite proof that $^{26}$O is indeed unbound was established in a recent invariant mass measurement at NSCL/MSU. The setup was similar to the previously discussed $^{16}$Be experiment. $^{26}$O was produced with the one-proton removal reaction from a secondary $^{27}$F beam. The three-body decay energy spectrum shown in Figure \ref{f:26O-data} shows a clear peak near threshold. The extracted decay energy was 150$^{+50}_{-150}$~keV \cite{Lun12}. The observed width was dominated by the experimental resolution so that the expected very narrow width of the ground state (see discussion below) could not be determined.

\begin{figure}[b]
\centerline{\includegraphics[height=.3\textheight]{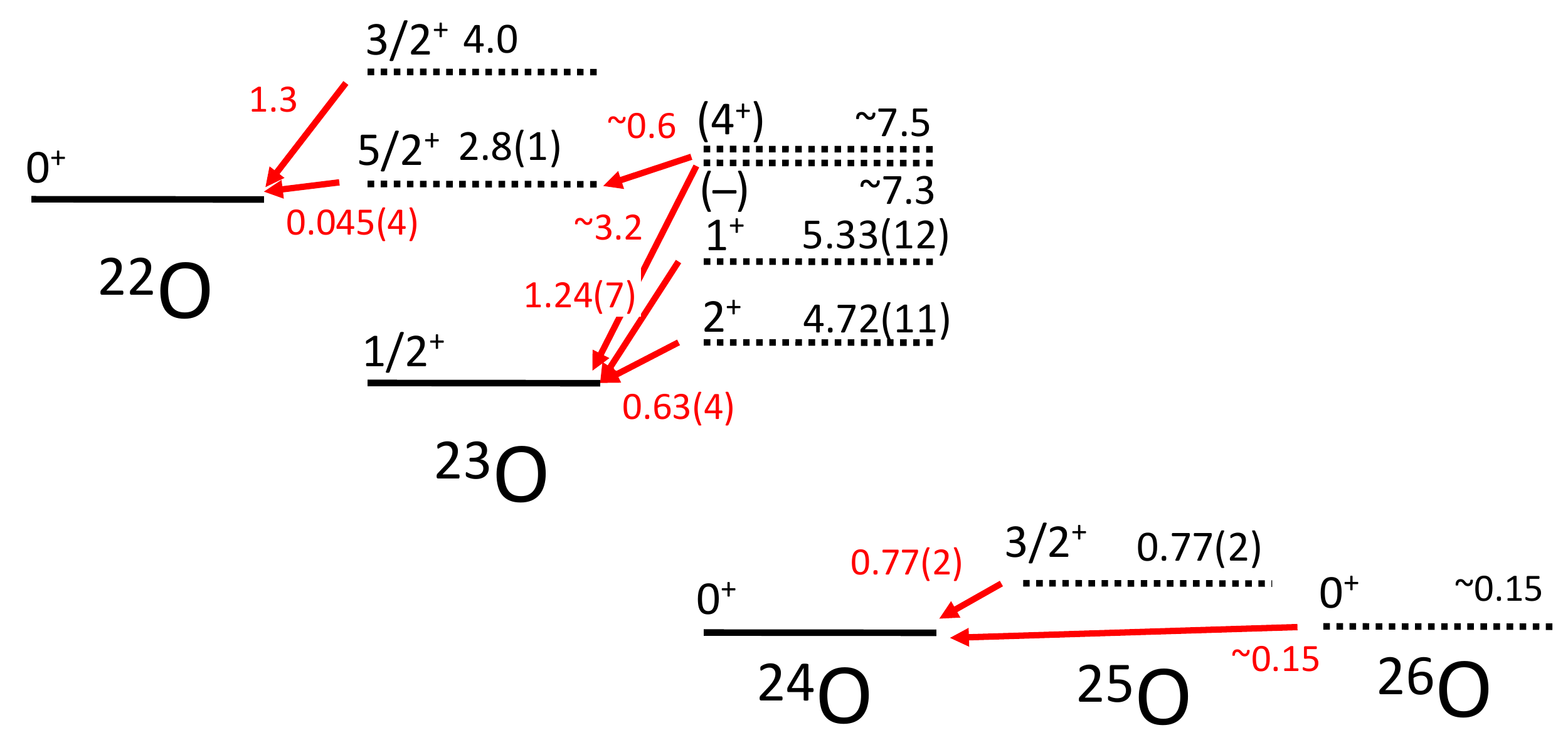}}
\caption{
Unbound levels in neutron-rich oxygen isotopes near and beyond the neutron dripline. All energies are in MeV and the decay energies are from \cite{Lun12,Hof09a,Ele07,Tsh12,Sch07,Hof09b,Hof11}.}
\label{f:Ox-level}
\end{figure}

The statistics were not sufficient to extract any correlations and thus the details of the decay could not be determined in this measurement. However, it is an even stronger candidate for a dineutron-like emission than $^{16}$Be, because the ground-state of the intermediate unbound system of $^{25}$O is unbound by 0.77(2)~MeV \cite{Hof09a} and thus is located about 600~keV above the $^{26}$O ground-state (See Figure \ref{f:Ox-level}). The figure also shows neutron-unbound states in the last two bound neutron-rich oxygen isotopes $^{23}$O and $^{24}$O. With the exception of the second excited state of $^{23}$O \cite{Ele07} and the negative parity state in $^{24}$O \cite{Tsh12} all measurements were performed with MoNA at NSCL/MSU \cite{Lun12,Hof09a,Sch07,Hof09b,Hof11}. The energies of the first two excited states of $^{24}$O were recently confirmed by the recent RIKEN measurement \cite{Tsh12}.

\begin{figure}
\centerline{\includegraphics[height=.4\textheight]{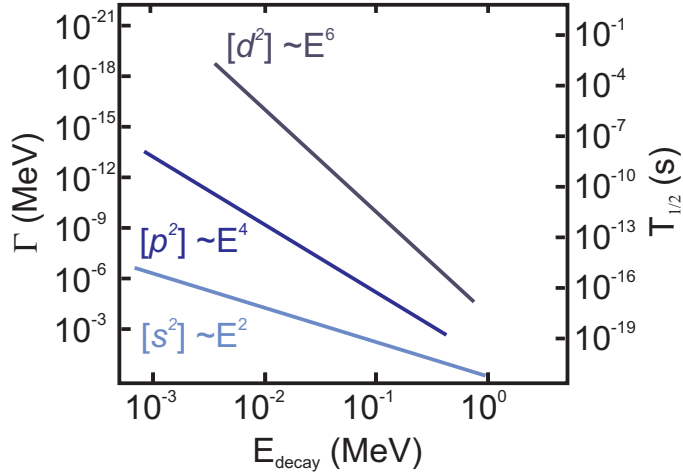}}
\caption{
Calculated widths and half-lives as a function of decay energy for two-neutron emission in different orbital configurations. (Adapted from Ref. \cite{Gri11})}
\label{f:grigorenko}
\end{figure}

The low decay energy of $^{26}$O very close to the two-neutron separation energy gives rise to speculations that $^{26}$O even represent a case for two-neutron radioactivity with a potentially fairly long lifetime. Grigorenko {\it el.} \cite{Gri11} have calculated the decay widths and half-lives for the two-neutron emission of $^{26}$O for different angular momenta of the neutrons (see Figure \ref{f:grigorenko}). The valence neutrons in $^{26}$O are most likely in a [$d^{2}$] configuration. A lower limit of 10$^{-14}$~s can be extracted from Figure \ref{f:grigorenko} for the $d^2$ configuration using the upper limit of the decay energy from the present experiment (200~keV). The unsuccessful searches for $^{26}$O using fragment separators yield an upper limit of the half-life of about 200~ns. Thus, the possible range of half-lives for $^{26}$O is still about 7 orders of magnitude with 2$\cdot$10$^{-7}$~s $<$ T$_{1/2} <$ 10$^{-14}$~s.

\section{Conclusion and Outlook}

Neutron decay spectroscopy with radioactive beams is an effective tool to explore nuclei at and beyond the neutron dripline. The two-neutron emission of $^{16}$Be corresponds to the first observation of a dineutron-like ground-state decay and the coincidence measurement of two-neutrons with $^{24}$O following the one-proton removal reaction from $^{27}$F determines unambiguously that $^{26}$O is unbound with respect to two-neutron emission.

The MoNA collaboration has recently commissioned an additional neutron detector array. LISA, the Large multi-Institutional Scintillator Array, was designed, built, and installed at the NSCL by a collaboration of undergraduate institutions \cite{Ste11}. The addition of LISA improves the efficiency, acceptance, and resolution for one- and especially two-neutron experiments. The data of the first experiment, where the decay of unbound excited states of $^{24}$O was measured with significantly improved resolution, are currently under analysis.

\begin{figure}
\centerline{\includegraphics[height=.4\textheight]{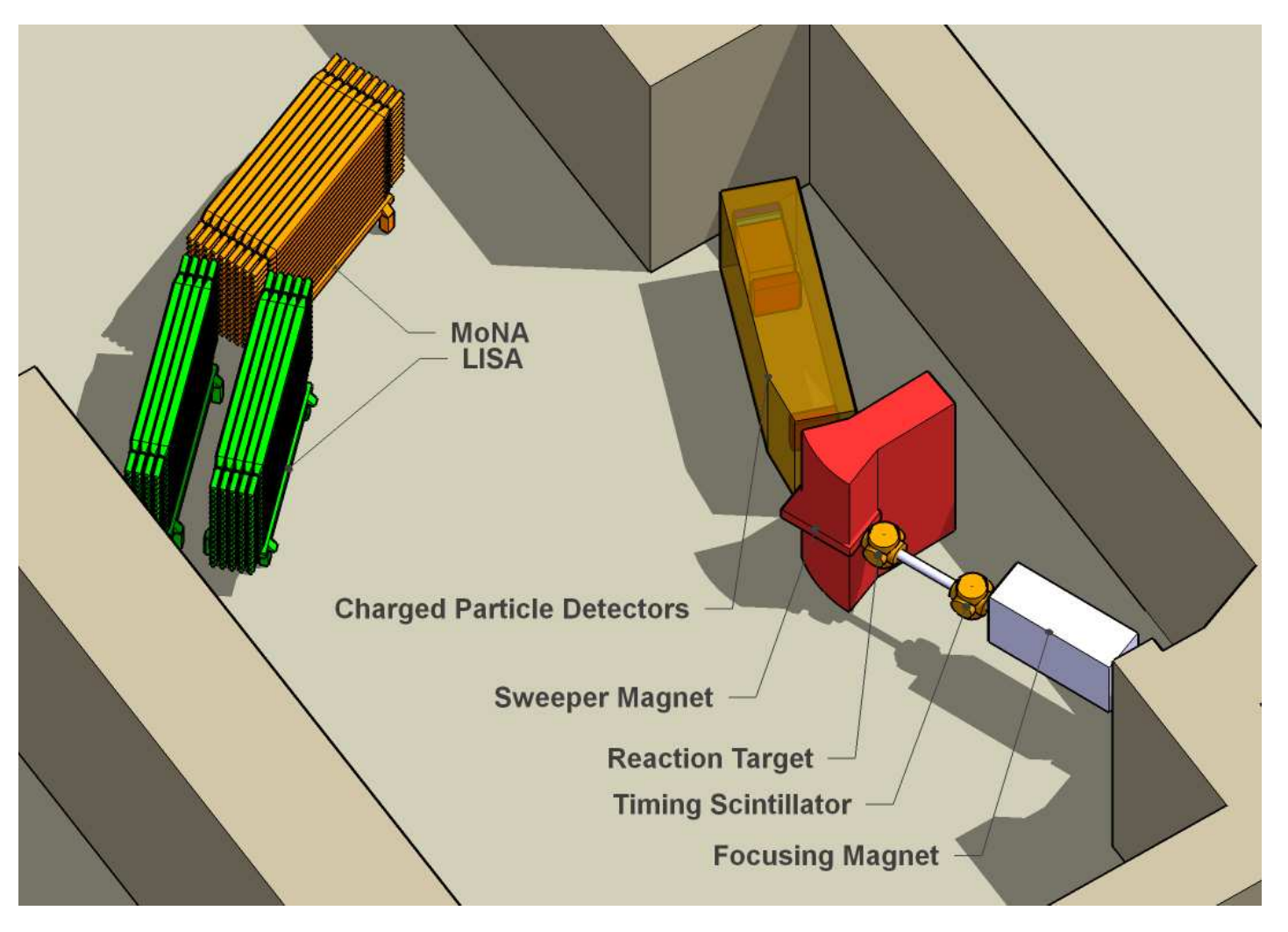}}
\caption{
Typical layout of the MoNA-LISA in at the NSCL. The beam enters the vault from the bottom right (Figure from \cite{Ste11}).}
\end{figure}

\section*{Acknowledgements}
This material is based upon work supported by the NSF under Grants PHY-02-44953, PHY-06-06007, PHY-07-57678, PHY-08-55456, PHY-09-69058, PHY-10-68217, and PHY-11-02511, and the DOE Awards DE-NA0000979 and DE-FG02-92ER40750.

\end{document}